# A Dynamics Perspective of Pursuit-Evasion Games of Intelligent Agents with the Ability to Learn

Hao Xiong, Huanhui Cao, Lin Zhang, and Wenjie Lu

*Abstract*—Pursuit-evasion games are ubiquitous in nature and in an artificial world. In nature, pursuer(s) and evader(s) are intelligent agents that can learn from experience, and dynamics (i.e., Newtonian or Lagrangian) is vital for the pursuer and the evader in some scenarios. To this end, this paper addresses the pursuit-evasion game of intelligent agents from the perspective of dynamics. A bio-inspired dynamics formulation of a pursuit-evasion game and baseline pursuit and evasion strategies are introduced at first. Then, reinforcement learning techniques are used to mimic the ability of intelligent agents to learn from experience. Based on the dynamics formulation and reinforcement learning techniques, the effects of improving both pursuit and evasion strategies based on experience on pursuit-evasion games are investigated at two levels: 1) individual runs and 2) ranges of the parameters of pursuit-evasion games. Results of the investigation are consistent with nature observations and the natural law – survival of the fittest. More importantly, with respect to the result of a pursuit-evasion game of agents with baseline strategies, this study achieves a different result. It is shown that, in a pursuit-evasion game with a dynamics formulation, an evader is not able to escape from a slightly faster pursuer with an effective learned pursuit strategy, based on agile maneuvers and an effective learned evasion strategy.

## I. INTRODUCTION

Pursuit-evasion game is a ubiquitous, fascinating, and valuable problem that has been studied in different disciplines (e.g., engineering, mathematics, and physics) in recent decades [1], and has still received lasting attention from different perspectives (e.g., optimal control [2], Homicidal Chauffeur problem [3], and game theory [4]). Achievements in pursuit-evasion games are valuable for a variety of applications, such as robot navigation [5], aircraft control [6], as well as self-driving vehicles [7]. A pursuit-evasion game includes two types of agents – pursuer(s) and evader(s). In a pursuit-evasion game, the pursuer chases the evader aiming to capture the evader, while the evader aims to avoid been captured. In nature, a prey and a predator can correspond to an evader and a pursuer, respectively [1]. In an artificial world, a pursuer and an evader can be mobile robots [8], vehicles [7], or Unmanned Aerial Vehicles (UAVs) [9].

A non-trivial scenario [1] in a pursuit-evasion game with a dynamics (i.e., Newtonian or Lagrangian) formulation is that a pursuer can run faster than its targeted evader (e.g., a cheetah, the ultimate cursorial pursuer, hunting a gazelle on open space, as shown in Fig.1, may use high velocity [10]) but has less agile maneuvers (i.e., the pursuer has a larger turning radius at higher velocity), while an evader can have more agile maneuvers (than the pursuer has) to possibly avoid capture [10]. In a pursuit-evasion game with a dynamics formulation, neither the pursuer nor the evader can adjust the magnitude or the direction of velocity to a target value instantaneously. An agent can only gradually change the magnitude or the direction of velocity through accelerating, decelerating, or turning, achieved by adjusting the driving force of itself. In the non-trivial scenario, effective pursuit and evasion strategies that take the driving force of an agent into account are vital. In recent decades, achievements of pursuit-evasion games have been reported from different perspectives, such as kinematics, motion-geometry, statistics, etc., without considering the driving forces of agents [1]. There are only a few researches on the pursuit-evasion game from the perspective of the dynamics considering the driving force.

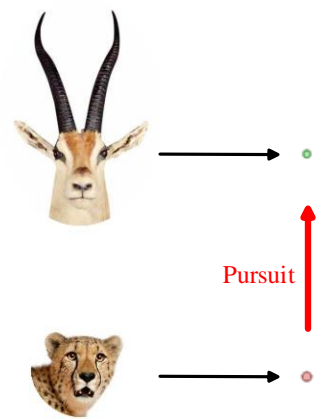

Fig. 1. A cheetah chases a gazelle in an open space

Li [1] conducted pioneering researches on a pursuit-evasion game from the perspective of dynamics. A general dynamics formulation of a pursuit-evasion game and baseline pursuit and evasion strategies were proposed, from a bio-inspired perspective, to approximate pursuit-evasion phenomena in nature. The concept of the *non-trivial escape zone* of a pursuit-evasion game concerning the maximum self-propelled acceleration of agents was proposed. The maximum

H. Xiong is with the School of Mechanical Engineering and Automation, Harbin Institute of Technology Shenzhen, Shenzhen, China (e-mail: xionghao@hit.edu.cn).

H. Cao is with the School of Mechanical Engineering and Automation, Harbin Institute of Technology Shenzhen, Shenzhen, China (e-mail: 20S153121@stu.hit.edu.cn).

L. Zhang is with the Department of Aerospace Engineering and Engineering Mechanics, University of Cincinnati, Cincinnati, OH 45040, USA (e-mail: lin.zhang@uc.edu).

W. Lu is with the School of Mechanical Engineering and Automation, Harbin Institute of Technology Shenzhen, Shenzhen, China (corresponding author: e-mail: luwenjie@hit.edu.cn).

self-propelled acceleration of agents included in the non-trivial escape zone indicates that the evader is able to escape from a faster pursuer, based on agile maneuvers and an effective evasion strategy.

Although Li's research [1] is innovative, the fact that the pursuer and the evader of a pursuit-evasion game in nature can be intelligent agents that can learn from their experience (denoted as intelligent agents in this paper) is not considered in [1]. It should be noticed that the pursuer and the evader of a pursuit-evasion game in nature are winners or witnesses of previous pursuit-evasion games and have accumulated experience. In view of the above-mentioned limit of [1], this paper considers the ability of intelligent agents to learn from experience, in addition to Li's research, and studies the effects of improving both pursuit and evasion strategies based on experience on pursuit-evasion games with dynamics formulations for the first time, to the authors' best knowledge.

To mimic the ability of intelligent agents to learn from the experience of pursuit-evasion games and improve strategies, reinforcement learning (RL) techniques are used in this paper. Via maximizing the expected cumulative rewards, RL investigates how an intelligent agent learns a strategy when the agent interacts with the environment [11]. RL has achieved great success in different areas (e.g., Atari [12], Go [13], and robot manipulation [14]). It is shown that RL can be used to address pursuit-evasion games with complex or unknown dynamics [15]. Wang et al. [15] have improved RL-based pursuit approaches for multi-player cooperative pursuit-evasion games via equipping pursuers with a communication mechanism. Asynchronous Advantage Actor-Critic (A3C) [16], an RL algorithm, was used to develop pursuit and evasion strategies for fixed-wing UAVs in [17]. A fuzzy deterministic policy gradient algorithm is introduced to address constant-velocity pursuit-evasion games in [18]. Zhu et al. [19] applied deep Q network (DQN) [12] algorithm to achieve evasion strategies of pursuit-evasion games as well. Successes of RL in [15], [17]–[19] verify the feasibility of applying RL to improve strategies of a pursuit-evasion game with a dynamics formulation in this study.

This paper studies pursuit-evasion games of intelligent agents with continuous state space and action space from the perspective of dynamics. The major contributions of this paper are as follows.

1) This paper applies RL to mimic the ability of intelligent agents of pursuit-evasion games to learn from the experience and studies the effects of this ability on pursuit-evasion games with dynamics formulations.

2) This paper shows that, in a pursuit-evasion game with a dynamics formulation, an evader is not able to escape from a slightly faster pursuer with an effective learned pursuit strategy, based on agile maneuvers and an effective learned evasion strategy.

The rest of the paper is organized as follows. Section II briefly introduces the dynamics formulation of a pursuit-evasion game, baseline pursuit and evasion strategies, and RL. Section III develops a method to simulate the ability of agents to learn from the experience of pursuit-evasion games with dynamics formulations and achieve learned strategies then. Section IV studies the effects of the ability of agents to learn from experience on individual runs of pursuit-evasion games. The effects of the ability of agents to learn from experience on the non-trivial escape zone are discussed in section V. Finally, section VI summarizes this paper.

## II. PRELIMINARIES

The preliminaries of the study of a pursuit-evasion game of intelligent agents with a dynamics formulation are demonstrated in this section, including the dynamics formation of a pursuit-evasion game, baseline pursuit and evasion strategies, and RL.

### A. Dynamics Formulation of a Pursuit-Evasion Game

In a predation pursuit-evasion scenario in nature, the pursuer approaches its targeted evader stealthy and suddenly runs toward the evader if the distance between the evader and the pursuer decreases to a threshold. Simultaneously the evader observes the pursuer and begins escaping. Based on this scenario, Li [1] proposed a dynamics formulation of a pursuit-evasion game and bio-inspired strategies to reveal natural laws. Let $x_p$ and $x_e$ denote the position of the pursuer and the evader, respectively. The dynamics formulation of the pursuer and the evader of the pursuit-evasion game can be expressed as

$$\ddot{x}_p(t) = -\mu \dot{x}_p(t) + a_p, \quad (1)$$

$$\ddot{x}_e(t) = -\mu \dot{x}_e(t) + a_e, \quad (2)$$

where $\dot{x}_p$ and $\ddot{x}_p$ denotes the velocity and acceleration of the pursuer. $\dot{x}_e$ and $\ddot{x}_e$ denotes the velocity and acceleration of the evader. $a_p$ and $a_e$ are the unit-mass self-propelled acceleration of the pursuer and the evader, respectively. $\mu$ is the velocity damping coefficient. $t$ represents the duration of the pursuit-evasion game.

### B. Baseline Pursuit and Evasion Strategies

To approximate pursuit-evasion phenomena in nature, bio-inspired baseline pursuit strategy $\pi_p^b$ and evasion strategy $\pi_e^b$ have been proposed for the pursuer and the evader as [1]

$$a_p = \pi_p^b\left(x_p(t), x_e(t)\right) = a_p d(t), \quad (3)$$

$$a_e = \pi_e^b(x_p(t), x_e(t), c)$$
$$= \begin{cases} a_e d(t), & \|x_p(t) - x_e(t)\| > c \\ a_e R(\dot{x}_p(t))d(t), & \|x_p(t) - x_e(t)\| \leq c \end{cases}, \quad (4)$$

where

$$d(t) = \frac{x_e(t) - x_p(t)}{\|x_p(t) - x_e(t)\|}. \quad (5)$$

$d(t)$ represents the unit vector from the pursuer to the evader. $R(\dot{x}_p(t))d(t)$ represents a normal vector of $d(t)$. $R(\dot{x}_p(t))d(t)$ indicates the suddenly turning propelling maneuvers. $c$ denotes a critical moment, at which the evader starts to perform turning propelling maneuvers. $\|*\|$ is the

Euclidean norm of $*$. $a_p$ and $a_e$ represent the maximum magnitude of the unit-mass self-propelled acceleration of the pursuer and the evader, respectively. Li [1] showed that within a certain range, a large value of $c$ favors the escape of the evader, while a too-large value of $c$ makes the evader hard to leave away from the pursuer. However, the optimal value of $c$ for the evader was not provided in [1].

*C. Reinforcement Learning*

A pursuit-evasion game can be formalized as a Markov Decision Process (MDP) [20]. The MDP can be defined by a tuple $\{S, A, p, r\}$ [21]. $S \in \mathbb{R}^n$ is the state space. $A \in \mathbb{R}^m$ is the action space. The transition probability $p: S \times A \times S \to [0,1]$ defines the dynamics of the MDP. The reward function $r: S \times A \to [r_{\min}, r_{\max}]$ provides the agent with a scalar reward at each state transition. A probability distribution $\pi: S \times A \to [0,1]$ represents the strategy. Typically, RL enables an agent to learn from experience and improve the strategy via maximizing the expected cumulative rewards $\sum_t \mathbb{E}_{\tau \sim \pi}[r(s_t, a_t)]$, where $\tau$ is the trajectory of an MDP episode. In recent year, various RL algorithms (e.g., DQN [12], deep deterministic policy gradient (DDPG) [14], and proximal policy optimization (PPO) [22]) and multi-agent RL algorithms (e.g., Neural Fictitious Self-Play (NFSP) [23], Policy-Space Response Oracles (PRSO) [24], multi-agent deep deterministic policy gradient (MADDPG) [25]) have been proposed to address RL problems.

## III. LEARNED STRATEGIES

With RL techniques, artificial agents in a simulation environment can mimic intelligent agents in nature to learn from experience. In this section, RL is applied to mimic the ability of intelligent agents to learn from experience in pursuit-evasion games. The pursuit-evasion game discussed in this study involves two agents with continuous state space and action space. Thus, the multi-agent deep deterministic policy gradient (MADDPG) algorithm [25] designed for multi-agent RL problems with continuous action space is used. Based on proper reward functions, MADDPG is suitable for problems with cooperation, competition, and a mixture of them, and has been used to study pursuit-evasion games in [15], [25]. Although multi-agent RL algorithms, such as MADDPG, suffer from the convergence issue [18], the convergence issue is not a problem in this study. Given the parameters of a pursuit-evasion game, one can adjust the hyper-parameters of MADDPG and rerun the pursuit-evasion game to achieve convergence eventually. The effectiveness of a learned pursuit strategy or a learned evasion strategy can be verified through playing the pursuit-evasion game with an adversarial baseline strategy.

The pursuit strategy $\pi_p$ and evasion strategy $\pi_e$ learned by intelligent agents can be expressed as

$$a_p = \pi_p\left(x_p(t), x_e(t), \dot{x}_p(t), \dot{x}_e(t)\right), \|a_p\| \le a_p, \quad (6)$$

$$a_e = \pi_e\left(x_p(t), x_e(t), \dot{x}_p(t), \dot{x}_e(t)\right), \|a_e\| \le a_e. \quad (7)$$

The strategies learned by intelligent agents and the baseline strategies are different mainly from two aspects. 1) Agents following the learned strategies do not have to perform their maximum magnitude of self-propelled acceleration, while agents following the baseline strategies always perform their maximum magnitude of self-propelled acceleration. 2) One has to set a critical moment for the baseline evasion strategy, while doesn't have to do so for the learned evasion strategy.

To achieve a simulation environment of a pursuit-evasion game with a dynamics formulation, the predator–prey environment developed by OpenAI [26] and used in [23, 38] is modified according to (1) and (2). The pursuit and evasion strategies of intelligent agents are learned based on the default hyper-parameters of MADDPG [25] in the modified simulation environment. In the pursuit-evasion game, the pursuer and the evader can observe the position and the velocity of both agents. The pursuer and the evader achieve the opposite reward as

$$r_e = -r_p = \begin{cases} 0.1\|x_p(t) - x_e(t)\|, & \|x_p(t) - x_e(t)\| > \varepsilon \\ -10, & \|x_p(t) - x_e(t)\| \le \varepsilon \end{cases}, \quad (8)$$

where $r_e$ and $r_p$ denote the reward of the evader and the pursuer, respectively. The evader is regarded as captured if the evader is within the capture radius $\varepsilon$ of the pursuer. The average reward in each episode of the pursuer with $a_p = 4$ and the evader with $a_e = 2.4$ in training are shown in Fig. 2. According to Fig. 2, the pursuit and evasion strategies converge in 1,700 episodes.

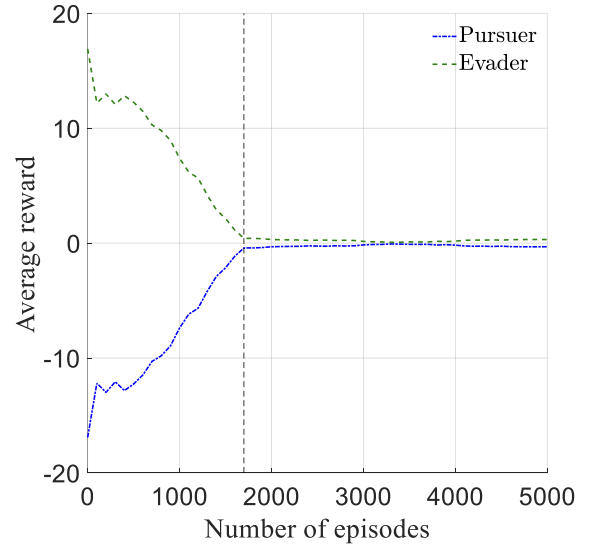

Fig. 2. Average rewards of the pursuer and the evader in training strategies

## IV. INDIVIDUAL RUNS

In this section, two cases of pursuit-evasion games discussed in [1] are investigated, considering that both the pursuer and the evader can learn from experience. It should be emphasized that when learning pursuit and evasion strategies, neither the pursuer nor the evader follows a corresponding baseline strategy. Thus, a learned strategy is not aiming to defeat the adversarial baseline strategy. The position and velocity of agents in the two cases are demonstrated in the former part of this section and the two cases are discussed in the latter part of this section.

The first case of pursuit-evasion games discussed in this section has the same setting as the pursuit-evasion game shown in Fig. 1 of [1]. In the first case, a pursuer with $a_p = 4$ and an evader with $a_e = 2$ and $c = 2.4$ play pursuit-evasion games in an environment with $\mu = 0.5$ and $\varepsilon = 0.5$. According to [1], the evader following the baseline evasion strategy $\pi_e^b$ cannot escape from the pursuer following the baseline pursuit strategy $\pi_p^b$. This scenario is replicated in this study, as shown in Fig. 3a. However, the evader following the learned evasion strategy $\pi_e$ can avoid been captured by the pursuer following the baseline pursuit strategy $\pi_p^b$, as shown in Fig. 3b. When the evader follows the learned evasion strategy $\pi_e$ and the pursuer follows the learned pursuit strategy $\pi_p$, it is shown that the evader cannot escape from the pursuer, as shown in Fig. 3c.

intelligent pursuer following the learned pursuit strategy $\pi_p$ levels off when it turns at 0.4 seconds, rather than increases continuously.

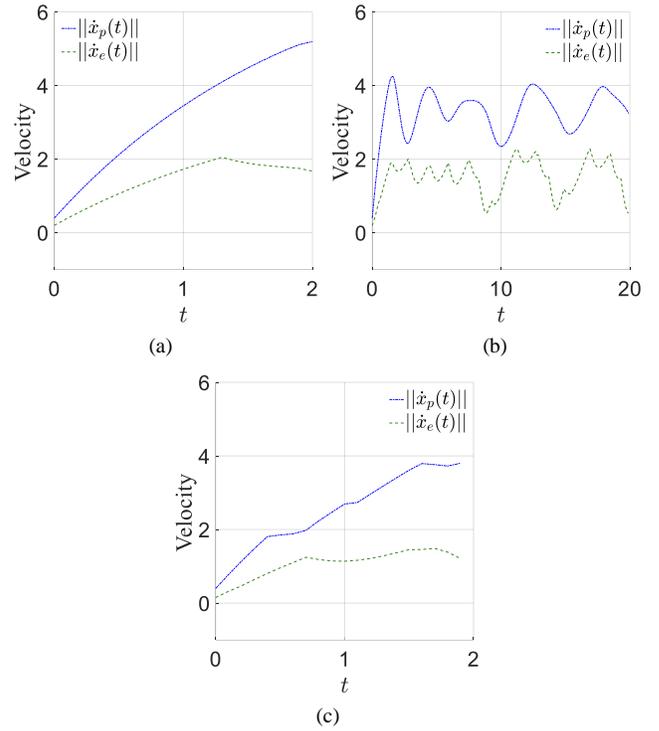

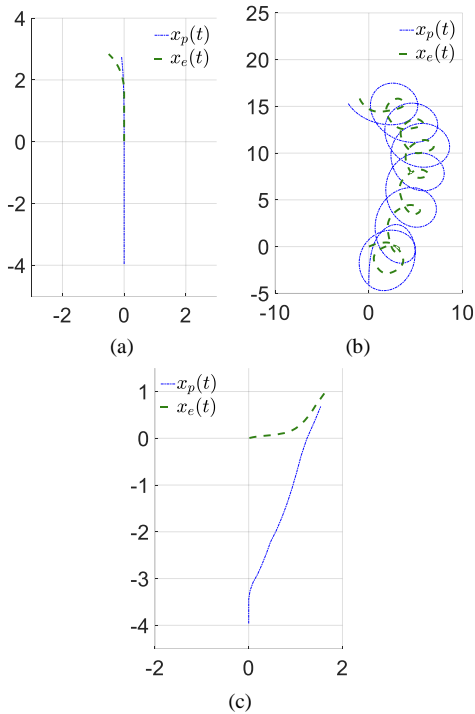

Fig. 3. $\mu = 0.5$, $a_p = 4$, $a_e = 2$, $c = 2.4$, $\varepsilon = 0.5$. Positions of agents with initial positions $\boldsymbol{x}_p(0) = (0,-4)$ and $\boldsymbol{x}_e(0) = (0,0)$, when a) the pursuer follows $\pi_p^b$ and the evader follows $\pi_e^b$ [1]; b) the pursuer follows $\pi_p^b$ and the evader follows $\pi_e$; c) the pursuer follows $\pi_p$ and the evader follows $\pi_e$

The velocities of the agents in the first case are present in Fig. 4. According to Fig. 4a, the pursuer following the baseline pursuit strategy $\pi_p^b$ accelerates continuously to catch the evader following the baseline evasion strategy $\pi_e^b$. Even though the evader tries to avoid capture via decreasing velocity to turn, the evader cannot avoid capture. Fig. 4b shows that the intelligent evader has learned to perform a sharp decrease in velocity just before turning, enabling tighter turns, as observed in nature [10]. As shown in Fig. 4c, the velocity of the

Fig. 4. $\mu = 0.5$, $a_p = 4$, $a_e = 2$, $c = 2.4$, $\varepsilon = 0.5$. Velocities of agents with initial positions $\boldsymbol{x}_p(0) = (0,-4)$ and $\boldsymbol{x}_e(0) = (0,0)$, when a) the pursuer follows $\pi_p^b$ and the evader follows $\pi_e^b$ [1]; b) the pursuer follows $\pi_p^b$ and the evader follows $\pi_e$; c) the pursuer follows $\pi_p$ and the evader follows $\pi_e$

The second case of pursuit-evasion games discussed in this section has the same setting as the pursuit-evasion game shown in Fig. 2 of [1]. In the second case, a pursuer with $a_p = 4$ and an evader with $a_e = 2.4$ and $c = 2.4$ play pursuit-evasion games in an environment with $\mu = 0.5$ and $\varepsilon = 0.5$. It is shown in [1] that the evader following the baseline evasion strategy $\pi_e^b$ can escape from the pursuer following the baseline pursuit strategy $\pi_p^b$. This study replicates this scenario, as shown in Fig. 5a. Nonetheless, the evader following the baseline evasion strategy $\pi_e^b$ cannot avoid been captured by the pursuer following the learned pursuit strategy $\pi_p$, as shown in Fig. 5b. Fig. 5c shows that when the evader follows the learned evasion strategy $\pi_e$ and the pursuer follows the learned pursuit strategy $\pi_p$, the evader still cannot escape from the pursuers. The difference between the two scenarios shown in Fig. 5b and Fig. 5c will be investigated in the next stage.

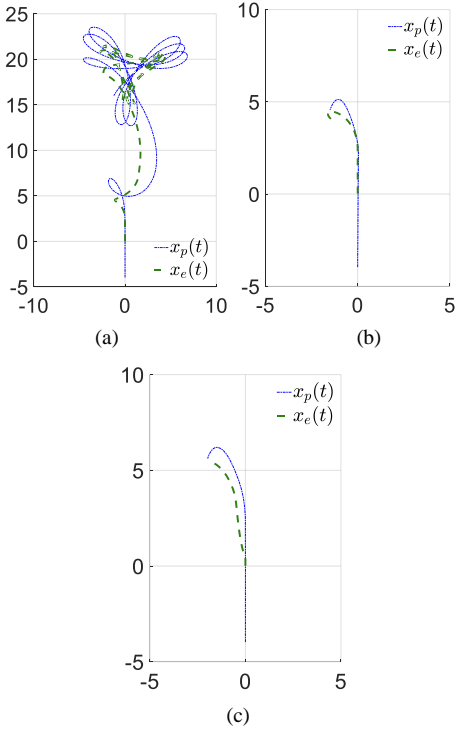

Fig. 5. $\mu = 0.5$, $a_p = 4$, $a_e = 2.4$, $c = 2.4$, $\varepsilon = 0.5$. Positions of agents with initial positions $\boldsymbol{x}_p(0) = (0, -4)$ and $\boldsymbol{x}_e(0) = (0,0)$, when a) the pursuer follows $\pi_p^b$ and the evader follows $\pi_e^b$ [1]; b) the pursuer follows $\pi_p$ and the evader follows $\pi_e^b$; c) the pursuer follows $\pi_p$ and the evader follows $\pi_e$

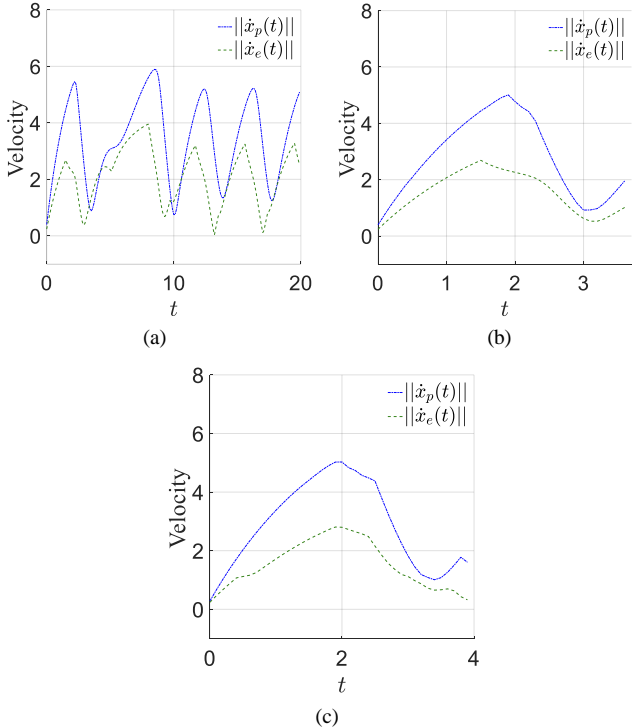

Fig. 6. $\mu = 0.5$, $a_p = 4$, $a_e = 2$, $c = 2.4$, $\varepsilon = 0.5$. Velocities of agents with initial positions $\boldsymbol{x}_p(0) = (0, -4)$ and $\boldsymbol{x}_e(0) = (0,0)$, when a) the pursuer follows $\pi_p^b$ and the evader follows $\pi_e^b$ [1]; b) the pursuer follows $\pi_p$ and the evader follows $\pi_e^b$; c) the pursuer follows $\pi_p$ and the evader follows $\pi_e$

The velocities of the agents in the second case are studied, as shown in Fig. 6. One can see from Fig. 6a that both the evader following the baseline evasion strategy $\pi_e^b$ and the pursuer following the baseline pursuit strategy $\pi_p^b$ can perform a sharp decrease in velocity just before turning, enabling tighter turns. Fig. 6b shows that when the evader follows the baseline evasion strategy $\pi_e^b$ and the pursuer follows the learned pursuit strategy $\pi_p$, the evader is captured in 3.7 seconds. While Fig. 6c presents that when the evader follows the learned evasion strategy $\pi_e$ and the pursuer follows the learned pursuit strategy $\pi_p$, the evader is captured in 4.0 seconds. This suggests that the intelligent evader following the learned evasion strategy $\pi_e$ can survive for a longer time, with respect to the evader following the baseline evasion strategy $\pi_e^b$. Also, it is shown in Fig. 6b and Fig. 6c that both the intelligent pursuer and the intelligent evader learn to decrease in velocity before turning, to perform tighter turns, which is consistent with the observations in nature [10]. The top velocity is not used by the intelligent pursuer following the learned pursuit strategy $\pi_p$ frequently when it is close to the evader, as observed in nature [10].

The results of the first and the second cases of pursuit-evasion games disclose some interesting effects of the ability of agents to learn from experience on pursuit-evasion games. According to Fig. 3b, a sophisticated intelligent evader can escape from a pursuer with the baseline pursuit strategy that can approximate pursuit-evasion phenomena in nature [1]. However, if the sophisticated intelligent evader suffers from another sophisticated intelligent pursuer, the pursuer can still capture the evader as shown in Fig. 3c. In nature, to survive in a pursuit-evasion game that is the same setting as the first case, the evader has to improve in aspects other than the evasion strategy (e.g., feasible self-propelled acceleration and utilizing terrain). Fig. 5b shows that a sophisticated intelligent pursuer can capture an evader following the baseline evasion strategy, even if a pursuer following the baseline pursuit strategy cannot capture the evader. This result indicates that an evader following the baseline evasion strategy may survive in several pursuit-evasion games in the beginning. However, without self-improvement, the evader may be captured by an intelligent pursuer later. One can see from Fig. 5c that if a pursuer with the advantage of self-propelled acceleration is intelligent and sophisticated, the pursuer can capture the intelligent evader, even if the sophisticated intelligent evader can learn from experience. For the evader, the improvement of the evasion strategy cannot make up for the weakness of its self-propelled acceleration.

The individual runs of the first and the second cases of pursuit-evasion games imply that baseline pursuit and evasion strategies that can approximate pursuit-evasion phenomena in nature [1] can be improved if agents can learn from experience. If a conservative agent that insists to follow a baseline strategy suffers from a sophisticated intelligent agent that improves its strategy continuously, the conservative agent may not win a pursuit-evasion game it could win in previous. The conservative agent will be selected out by competition, according to natural selection. This phenomenon is consistent with the natural law - survival of the fittest, and shows the merit of considering the ability of agents to learn from experience in studies of pursuit-evasion games in nature.

## V. NON-TRIVIAL ESCAPE ZONE

The effects of the ability of agents to learn from experience on the non-trivial escape zone of pursuit-evasion games proposed in [1] are investigated in this section. The non-trivial escape zone concerns the maximum self-propelled acceleration of the pursuer and the evader. If the maximum self-propelled acceleration of the pursuer and the evader is included in the non-trivial escape zone, the evader is able to escape from a faster pursuer, based on agile maneuvers and an effective evasion strategy. The escape zone of pursuit-evasion games of intelligent agents following learned pursuit and evasion strategies is achieved, as shown in Fig. 7b. This escape zone is compared to the escape zone of pursuit-evasion games of agents following baseline pursuit and evasion strategies obtained in [1], as shown in Fig. 7a.

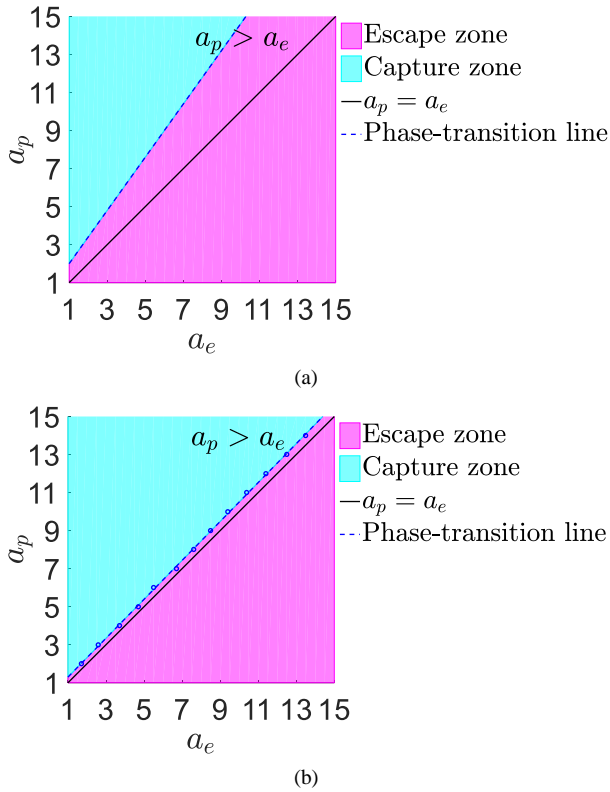

Fig. 7. $\mu = 0.5$, $\varepsilon = 0.5$, $t \leq 20$ seconds. Capture zone and escape zone when agents have initial positions $x_p(0) = (0, -12)$ and $x_e(0) = (0,0)$ and a) adopt baseline strategies $\pi_p^b$ or $\pi_e^b$ ( $c = 3$ ) correspondingly [1]; b) adopt learned strategies $\pi_p$ or $\pi_e$ correspondingly

In Fig. 7, the non-trivial escape zone refers to the magenta-color zone with $a_p > a_e$. Given pursuit and evasion strategies, there is a phase-transition line, which separates the capture zone and the escape zone of pursuit-evasion games. In Fig. 7a, the phase-transition line of pursuit-evasion games of agents following the baseline pursuit and evasion strategies can be expressed as [1]

$$a_p \approx 1.4a_e + 0.6. \qquad (9)$$

For pursuit-evasion games of intelligent agents, the phase-transition line can be approximated as

$$a_p \approx 1.023a_e + 0.275. \qquad (10)$$

One can find a large non-trivial escape zone from Fig. 7a. The large non-trivial escape zone in Fig. 7a suggests that in a pursuit-evasion game with a dynamics formulation, an evader can escape from a much faster pursuer with a baseline pursuit strategy, based on agile maneuvers and a baseline evasion strategy. Fig. 7b shows that if both agents of a pursuit-evasion game have the ability to learn from experience, the area of the non-trivial escape zone is much smaller than that of a pursuit-evasion game of agents following the baseline pursuit and evasion strategies. One can see from Fig. 7 that although an intelligent evader can utilize agile maneuvers and an effective learned evasion strategy to escape from a faster pursuer, the advantage brought by agile maneuvers is limited if both the evader and the pursuer are intelligent and sophisticated. In a pursuit-evasion game with a dynamics formulation, an evader is not able to escape from a slightly faster pursuer with an effective learned pursuit strategy, based on agile maneuvers and an effective learned evasion strategy. This result is different from the result of a pursuit-evasion game of agents following baseline strategies. This result also suggests that, in addition to an effective evasion strategy, self-propelled acceleration is vital for the evader, if the pursuer can learn from its experience.

## VI. CONCLUSION

This paper addressed a pursuit-evasion game of intelligent agents from the perspective of dynamics. Base on RL techniques and a bio-inspired dynamics formulation of a pursuit-evasion game, the effects of improving both pursuit and evasion strategies by making use of experience on pursuit-evasion games were studied. This study achieved experiment results that are consistent with natural observations (e.g., a sharp decrease in velocity before turning, enabling tighter turns) and the natural law - survival of the fittest. It is shown that, in a pursuit-evasion game with a dynamics formulation, an evader is not able to escape from a slightly faster pursuer with an effective learned pursuit strategy, based on agile maneuvers and an effective learned evasion strategy.

This study also has some limitations. Firstly, the learned pursuit and evasion strategies may be over-fitted to the pursuit-evasion game included in training. Secondly, the time delay issue of pursuit-evasion games in nature has not been considered. These limitations will be addressed in our future researches.